\documentclass[floatfix, nofootinbib]{revtex4}
\usepackage{graphicx}
\usepackage{amsmath}

\begin{document}

\title{Collision Energy Evolution of Elliptic and Triangular Flow in a Hybrid Model\footnote{
Talk given in CPOD 2013 - 8th International Workshop on Critical Point and Onset of Deconfinement,
March 11-15, Napa, CA}}

\author{J.~Auvinen\footnote{Speaker.}}
\email{auvinen@fias.uni-frankfurt.de}

\author{H.~Petersen}
\email{petersen@fias.uni-frankfurt.de}
\affiliation{Frankfurt Institute for Advanced Studies (FIAS), Ruth-Moufang-Strasse 1, D-60438 Frankfurt am Main, Germany}

\begin{abstract}
While the existence of a strongly interacting state of matter, known as
``quark-gluon plasma'' (QGP), has been established in heavy ion collision 
experiments in the past decade, the task remains to map out the transition from 
the hadronic matter to the QGP. This is done by measuring the dependence of key 
observables (such as particle suppression and elliptic flow) on the collision energy
of the heavy ions. This procedure, known as "beam energy scan", has been
most recently performed at the Relativistic Heavy Ion Collider (RHIC).

Utilizing a Boltzmann+hydrodynamics hybrid model, we study
the collision energy dependence of initial state eccentricities and the
final state elliptic and triangular flow. This approach is well suited
to investigate the relative importance of hydrodynamics and hadron
transport at different collision energies. 
\end{abstract}

\maketitle

\section{Introduction}

The RHIC beam energy scan program was launched in 2010 to study the features of 
the QCD phase diagram. In particular, the goal is to search for signs of the possible 
1st-order phase transition between the confined and deconfined matter, and 
locate the critical point marking the boundary of cross-over and 1st-order phase 
transition in the plane of baryochemical potential $\mu_B$ and temperature $T$ 
\cite{Kumar:2011de}, predicted by lattice calculations 
\cite{cite_lattice}.

Elliptic flow is one of the key observables that supports the finding of a strongly 
coupled quark-gluon plasma at the highest energies of RHIC and the Large Hadron Collider 
(LHC). Therefore, one would naively expect the elliptic flow to decrease at lower beam 
energies where the hydrodynamic phase gets shortened or the QGP is not 
created at all. It has been found, however, that the inclusive charged hadron elliptic 
flow $v_2$ demonstrates very little dependence on the collision energy between 
$\sqrt{s_{NN}}=7.7 - 39$ GeV \cite{Adamczyk:2012ku}. 

The beam energy dependence of the collective flow has been recently studied with several 
different models \cite{cite_models}. 
One possible method for investigating the importance of the hydrodynamical evolution for the 
flow production is the hybrid approach, where a transport model 
(a microscopic description of the system) is utilized for the non-equilibrium phases at 
the beginning and the end of a heavy-ion collision event, and a (macroscopic) 
hydrodynamical description is used to model the hot and dense stage and the phase 
transition between the QGP and hadronic matter.

As such a hybrid model should be able to naturally produce the transition from the 
high-energy heavy ion collisions, with negligible net-baryon density and a large 
hydrodynamically evolving medium, to smaller energies with finite net-baryon density 
and lower temperatures, where no such medium is formed, this framework seems optimal 
for studying the beam energy dependence of the elliptic and triangular flow.

\section{Hybrid model}

This study was performed using a transport + hydrodynamics hybrid model described 
in \cite{Petersen:2008dd}. In this approach, the initial state is produced by the 
Ultrarelativistic Quantum Molecular Dynamics (UrQMD) string / hadronic cascade 
\cite{cite_urqmd}.
The hydrodynamical evolution starts, when the two colliding nuclei have passed through 
each other: $t_{\textrm{start}}=\textrm{max}\{\frac{2R}{\sqrt{\gamma_{CM}^2-1}},0.5 \textrm{ fm}\}$, 
where $R$ represents the nuclear radius and $\gamma_{CM}=\frac{1}{\sqrt{1-v_{CM}^2}}$ 
is the Lorentz factor. The minimum time of 0.5 fm is chosen based on the hybrid model 
results at the collision energy $\sqrt{s_{NN}}=200$ GeV \cite{Petersen:2010zt}.
At this time, the energy-, momentum- and baryon number densities of the particles, 
represented by 3D Gaussian distributions that are Lorentz-contracted in the beam direction, 
are mapped onto the hydro grid. The width parameter of these Gaussians is $\sigma=1.0$ fm, 
to preserve the event-by-event initial state fluctuations. Spectators do not participate on 
the hydrodynamical evolution, but are propagated separately in the cascade.

The model utilizes (3+1)-D ideal hydrodynamics, solving the evolution equations using 
the SHASTA algorithm \cite{cite_shasta}. The equation of state is 
based on a chiral model, coupled to Polyakov loop to include the deconfinement phase 
transition \cite{Steinheimer:2009nn,Steinheimer:2010ib}, which qualitatively agrees with 
the lattice QCD data at $\mu_B=0$ and is also applicable at finite baryon densities. After 
the last step of the hydrodynamical evolution, the active EoS is changed from the 
deconfinement EoS to the hadron gas EoS, to ensure that the active degrees of freedom 
on both sides of the transition hypersurface are exactly equivalent \cite{Steinheimer:2009nn}.

The transition from hydro to transport (``particlization'') is done 
when the energy density $\epsilon$ is smaller than the critical value $2\epsilon_0$,
where $\epsilon_0=146$ MeV/fm$^3$ represents the nuclear ground state energy density. 
This corresponds roughly to a switching temperature $T \approx 154$ MeV at 
$\sqrt{s_{NN}}=200$ GeV Au+Au collisions \cite{Huovinen:2012is}. The switching criterion 
with respect to the energy density is kept constant over all beam energies in this 
study, but naturally corresponds to different combinations of temperature and 
baryochemical potential at different values of $\sqrt{s_{NN}}$.

From the iso-energy density hypersurface, constructed using the Cornelius algorithm 
\cite{Huovinen:2012is}, particle distributions are generated according to the Cooper-Frye 
formula. After the particlization, rescatterings and final decays are computed in the UrQMD. 
The end result is a distribution of particles which can be directly compared against 
the experimental data.

This hybrid approach has the advantage of dynamically changing the importance of the 
non-equilibrium transport and the hydrodynamic part of the evolution and involves 
a proper equation of state that is applicable at high net baryon densities. 
The high viscosity during the hadron gas evolution is taken into account, while the 
small viscosity during the hydrodynamic evolution has been neglected for simplicity 
to demonstrate qualitative behavior. 

\section{Results}

\subsection{Particle spectra}

\begin{figure}
\centering
\includegraphics[width=7.4cm]{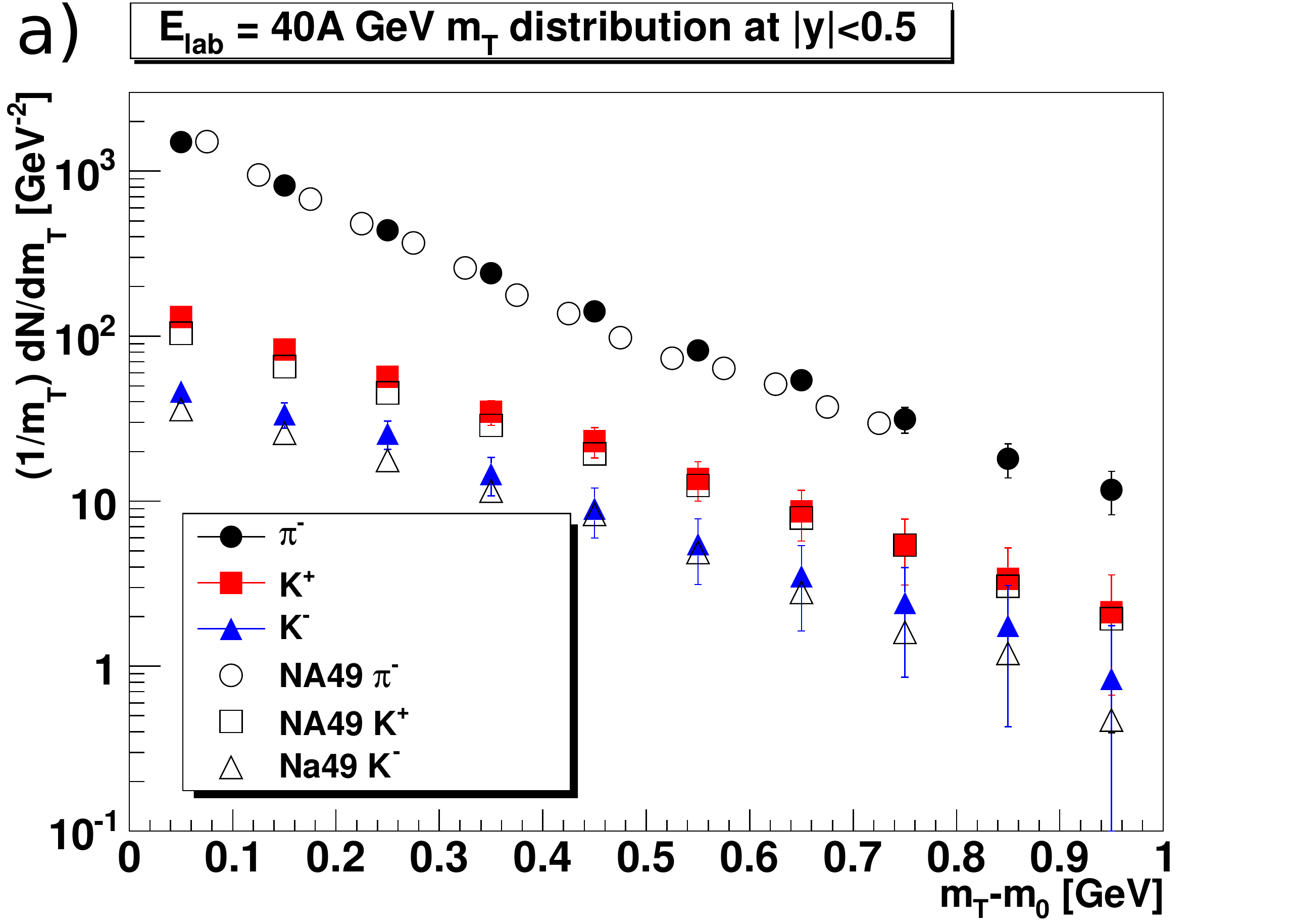}
\includegraphics[width=7.4cm]{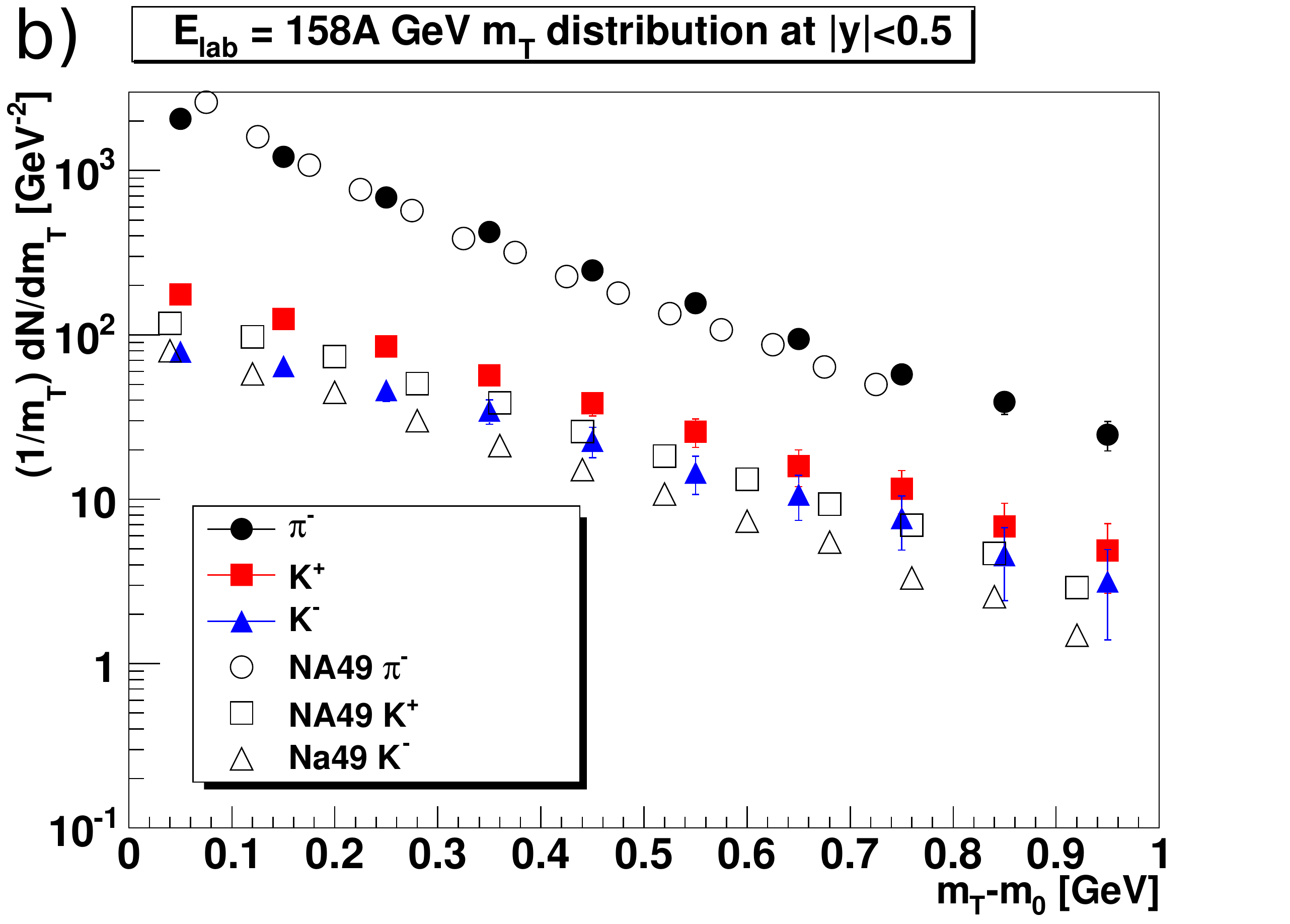}
\caption{Transverse mass spectra at midrapidity $|y|<0.5$ for $\pi^-, K^+$ and $K^-$ in 
Pb+Pb -collisions, compared to the NA49 data \cite{Afanasiev:2002mx} at beam energy 
a) $E_{\textrm{lab}}=40$ AGeV, b) $E_{\textrm{lab}}=158$ AGeV.}
\label{Figure_mt_spectra}
\end{figure}

Before going into more detailed observables such as the elliptic flow, we check how well 
the hybrid model reproduces the more general features of the system, such as the particle 
spectra. The evolution of $m_T$ spectra at midrapidity $|y|<0.5$ for $\pi^-, K^+$ and $K^-$ 
as a function of beam energy in Pb+Pb -collisions is illustrated in 
Figure~\ref{Figure_mt_spectra}. For the beam energy $E_{\textrm{lab}}=40$ AGeV, 
corresponding to the collision energy $\sqrt{s_{NN}}\approx 9$ GeV, there is a good 
agreement with the NA49 data \cite{Afanasiev:2002mx}. However, at the higher energies it 
becomes clear that the pion slope is a little too flat and there is an excess of kaons 
produced. This necessitates revisiting the model parameters, chiefly the value of 
particlization energy density, in the future studies. Nevertheless, for the purpose of the 
current investigation the agreement with the experimental data is sufficient.

\subsection{Elliptic flow}

Our primary interest here is to see, if the insensitivity of the elliptic flow $v_2$ 
on the collision energy can be understood within the hybrid approach. In this study $v_2$ is 
computed from the particle momentum distributions using the event plane method 
\cite{cite_eventplane}. This and the new implementation of 
the Cooper-Frye hypersurface finder and particlization are the main differences 
in this calculation compared to previous studies of elliptic flow 
in the same hybrid approach \cite{cite_hybrid}.  

\begin{figure}
\centering
\includegraphics[width=8cm]{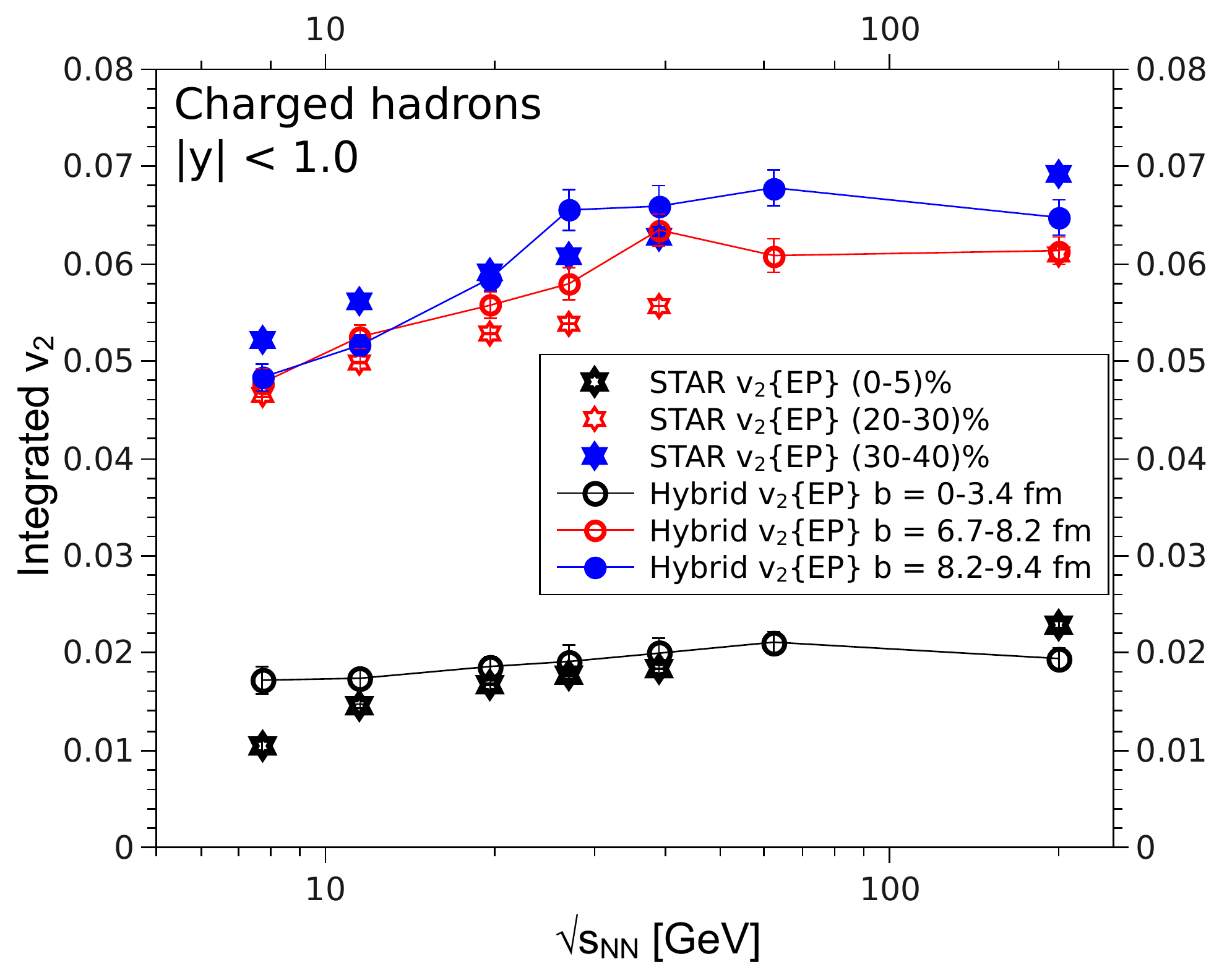}	
\caption{Integrated elliptic flow $v_2$ at midrapidity $|y|<1.0$ in Au+Au -collisions, 
for collision energies $\sqrt{s_{NN}}=7.7 - 200$ GeV and three different impact parameter 
ranges, compared with the STAR data \cite{Adamczyk:2012ku,Adams:2004bi}.}
\label{Figure_v2_star}
\end{figure}

Figure~\ref{Figure_v2_star} shows the produced $p_T$-integrated elliptic flow $v_2$ in 
Au+Au -collisions, compared with the STAR data for three centrality 
classes: (0-5)\%, (20-30)\% and (30-40)\%. In the model these are respectively represented 
by the impact parameter intervals $b = 0-3.4$ fm, $b = 6.7-8.2$ fm and $b = 8.2-9.4$ fm, 
based on the optical Glauber model estimates \cite{Eskola:1988yh}.

The agreement with the experimental data in the most central collisions is good above 
$\sqrt{s_{NN}}=11.5$ GeV; at the lowest energies the model appears to produce more flow 
than is observed in the experiments. However, in midcentral collisions the hybrid model 
does reproduce the measured increase of $v_2$ with respect to $\sqrt{s_{NN}}$.

\begin{figure}
\centering
\includegraphics[width=7.6cm]{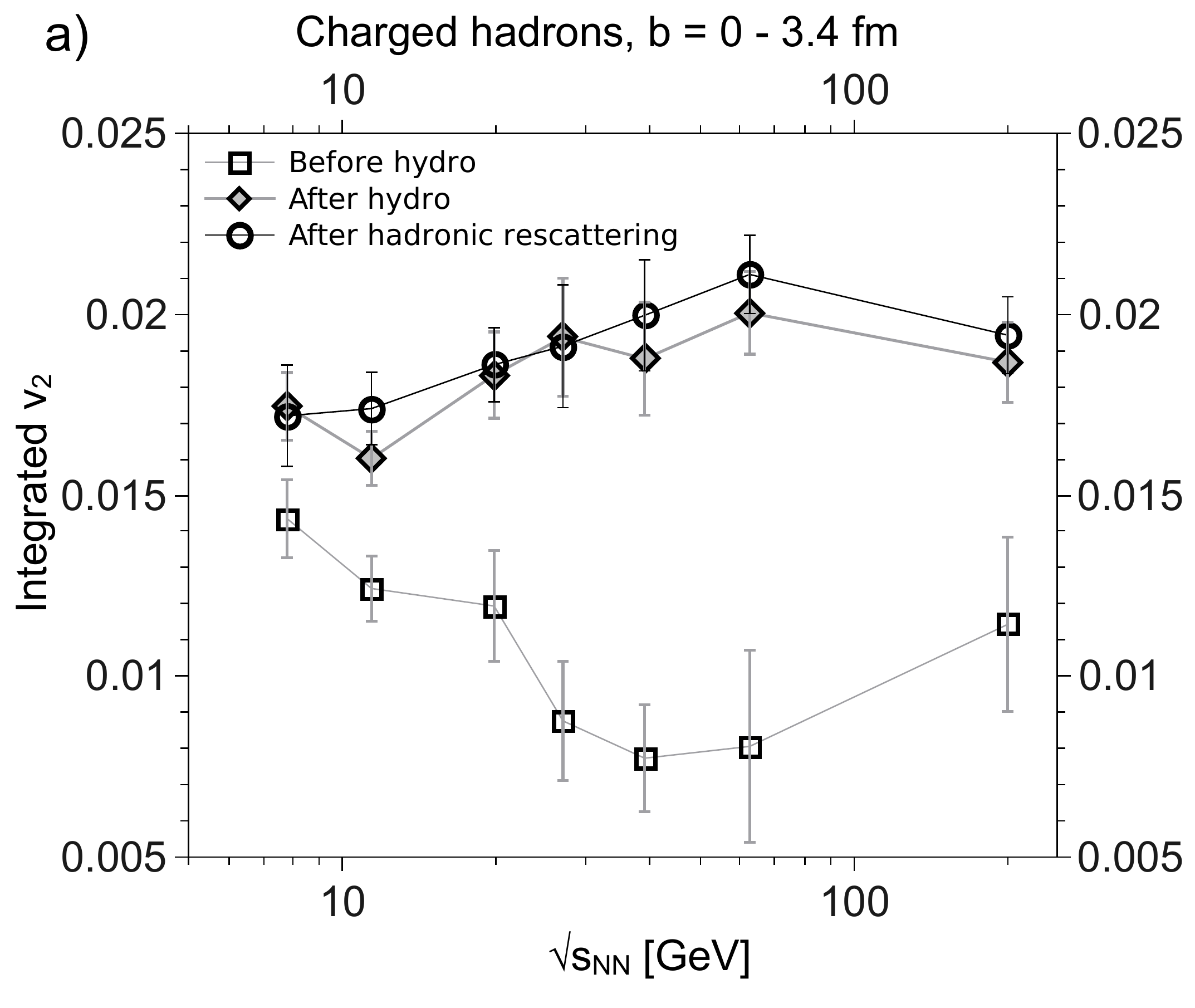}
\includegraphics[width=7.3cm]{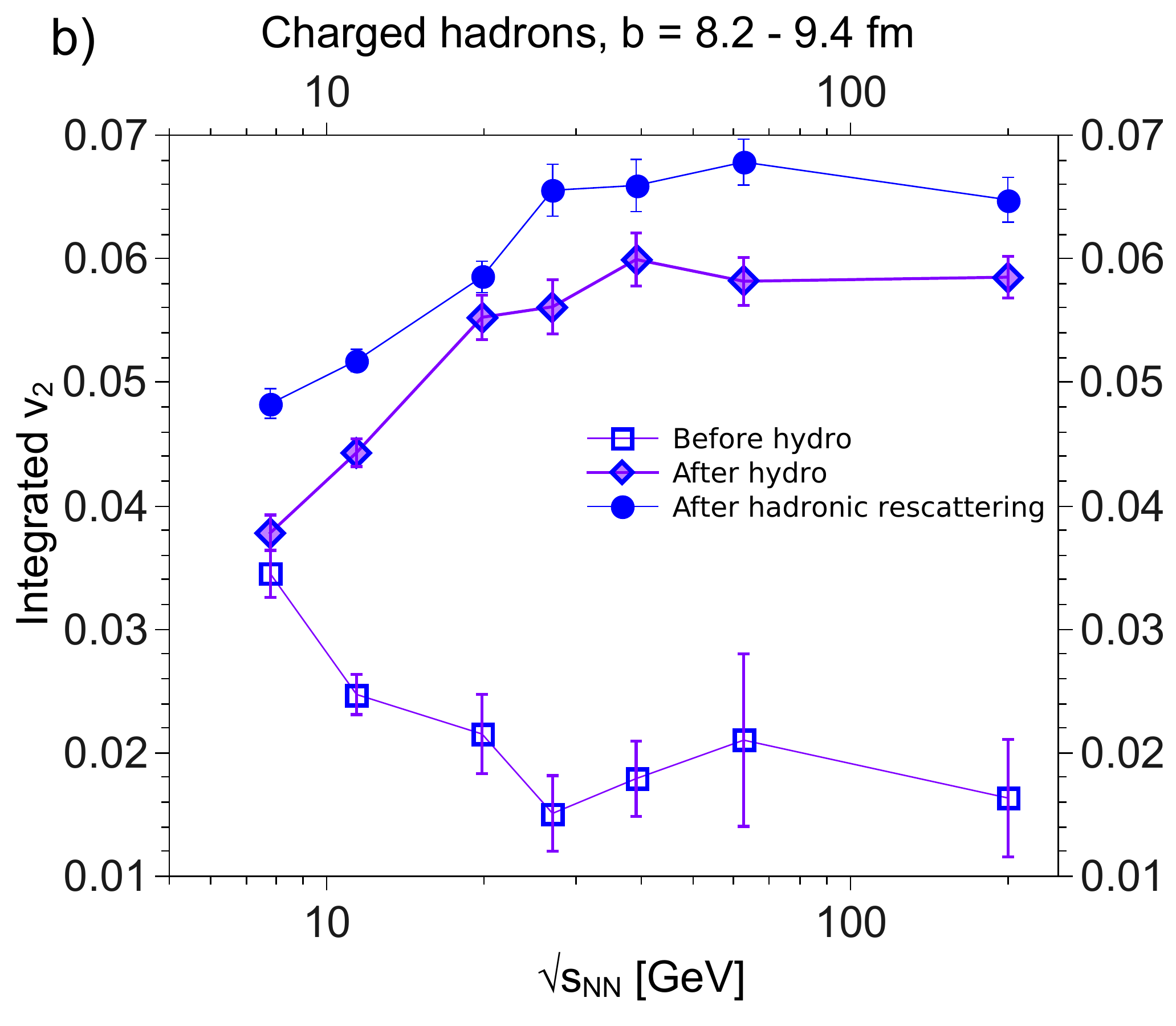}
\caption{Magnitude of $v_2$ at the beginning of hydrodynamical evolution (squares),
immediately after particlization (diamonds) and after the full simulation (circles, the same
as in Fig.~\protect\ref{Figure_v2_star}) at a) central collisions and b) midcentral 
collisions.}
\label{Figure_v2_phases}
\end{figure}

We can now investigate in more detail the contribution to $v_2$ from different phases 
of the heavy ion collision event. Figure~\ref{Figure_v2_phases} demonstrates the magnitude 
of $v_2$ before the hydrodynamical evolution, right after particlization and 
finally after the hadronic rescatterings performed in the UrQMD (the end result). 
In the most central collisions, where the overall elliptic flow is small compared to 
mid-central collisions, the effect of the hadronic rescatterings is negligible. In 
the impact parameter range $b = 8.2-9.4$ fm the contribution from the hadronic 
rescatterings is about 10\%. 

In both centralities, it is observed that at $\sqrt{s_{NN}}=7.7$ GeV, hydrodynamics 
contribute very little to the elliptic flow; for the mid-central collisions, 
$v_2$ is in practice completely produced by the transport dynamics. However, already at 
$\sqrt{s_{NN}}=11.5$ GeV the contribution from the hydrodynamic phase is significant.

\begin{figure}
\centering
\includegraphics[width=4.7cm]{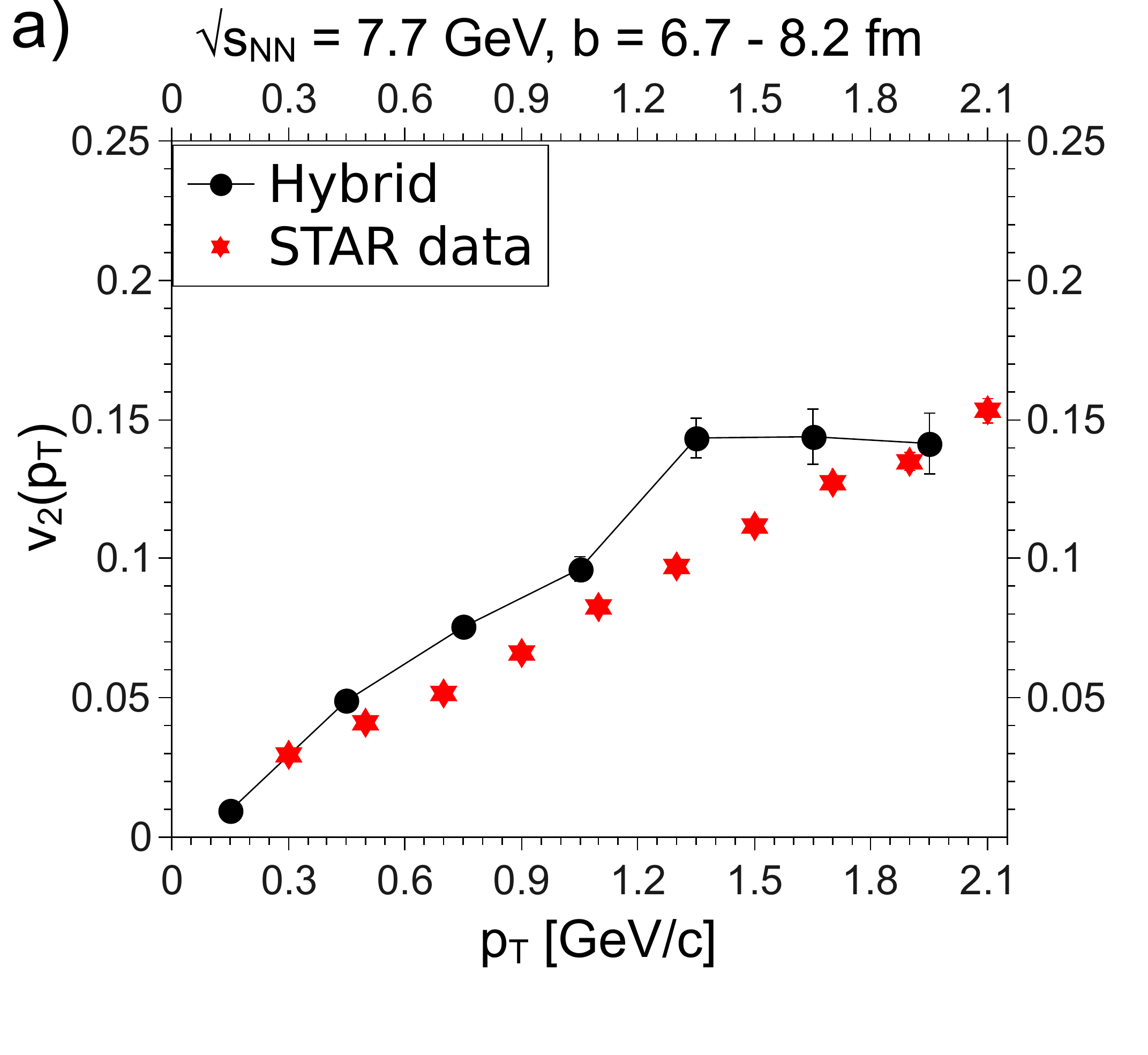}
\includegraphics[width=4.7cm]{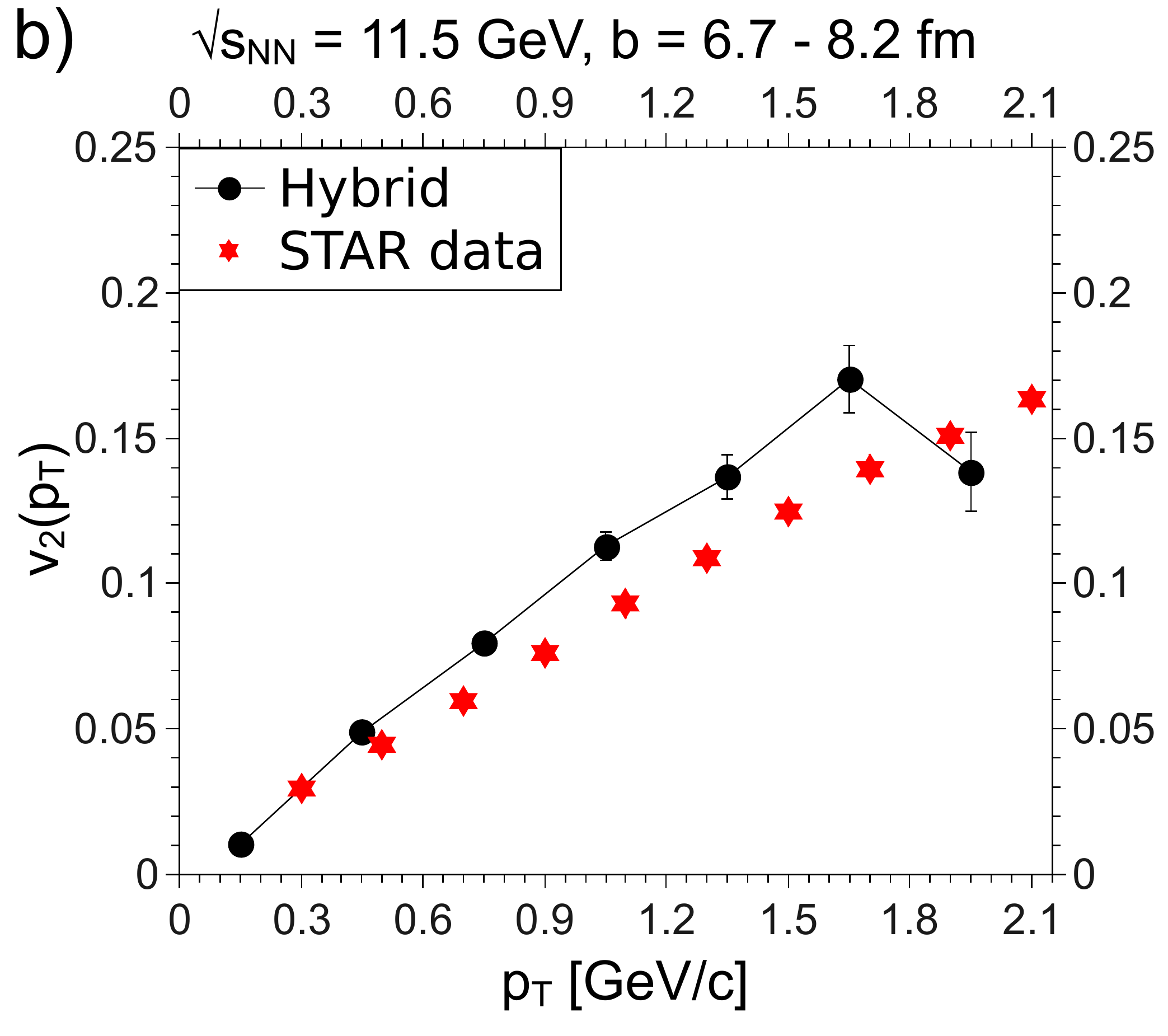}
\includegraphics[width=4.7cm]{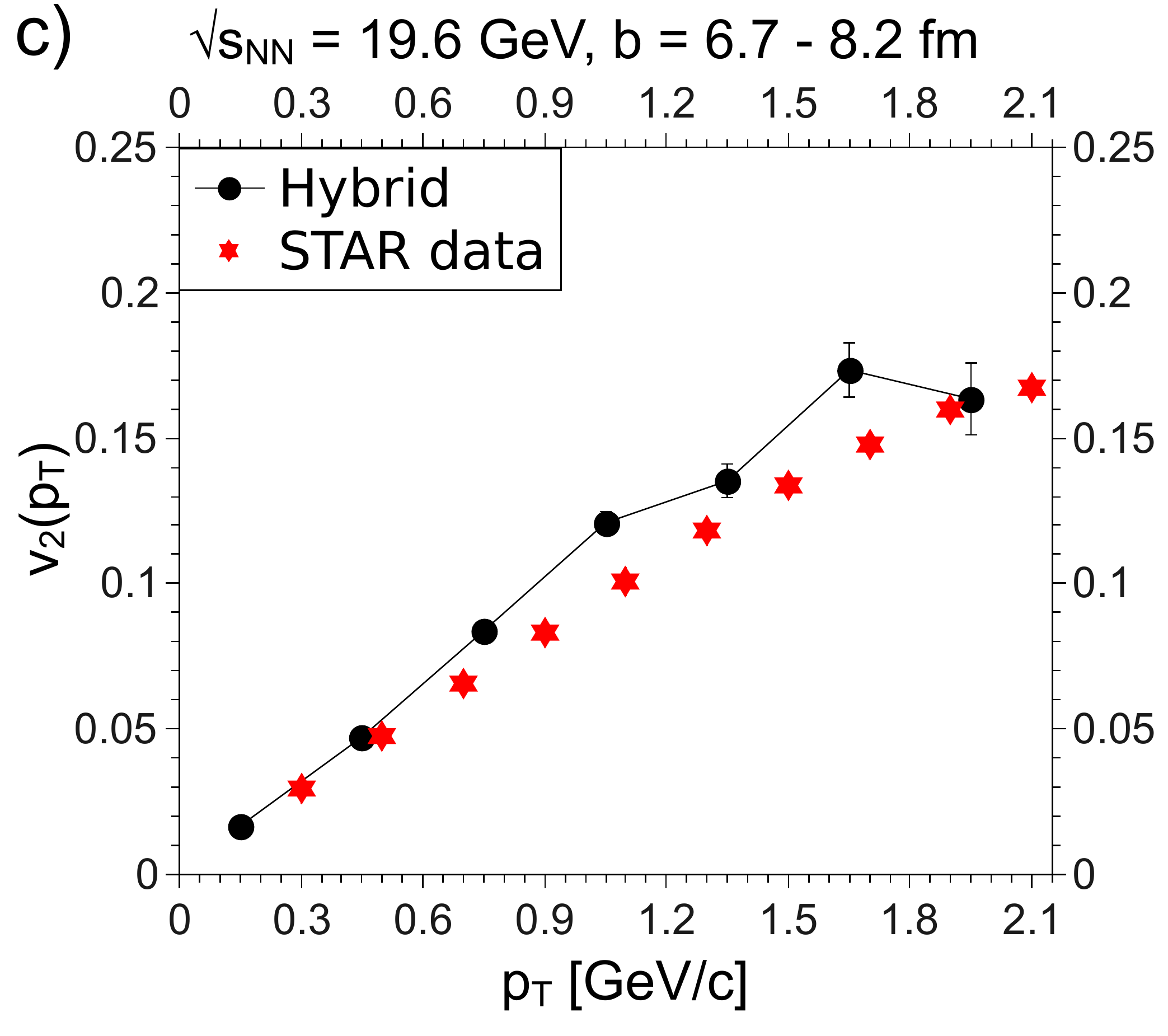}	
\includegraphics[width=4.7cm]{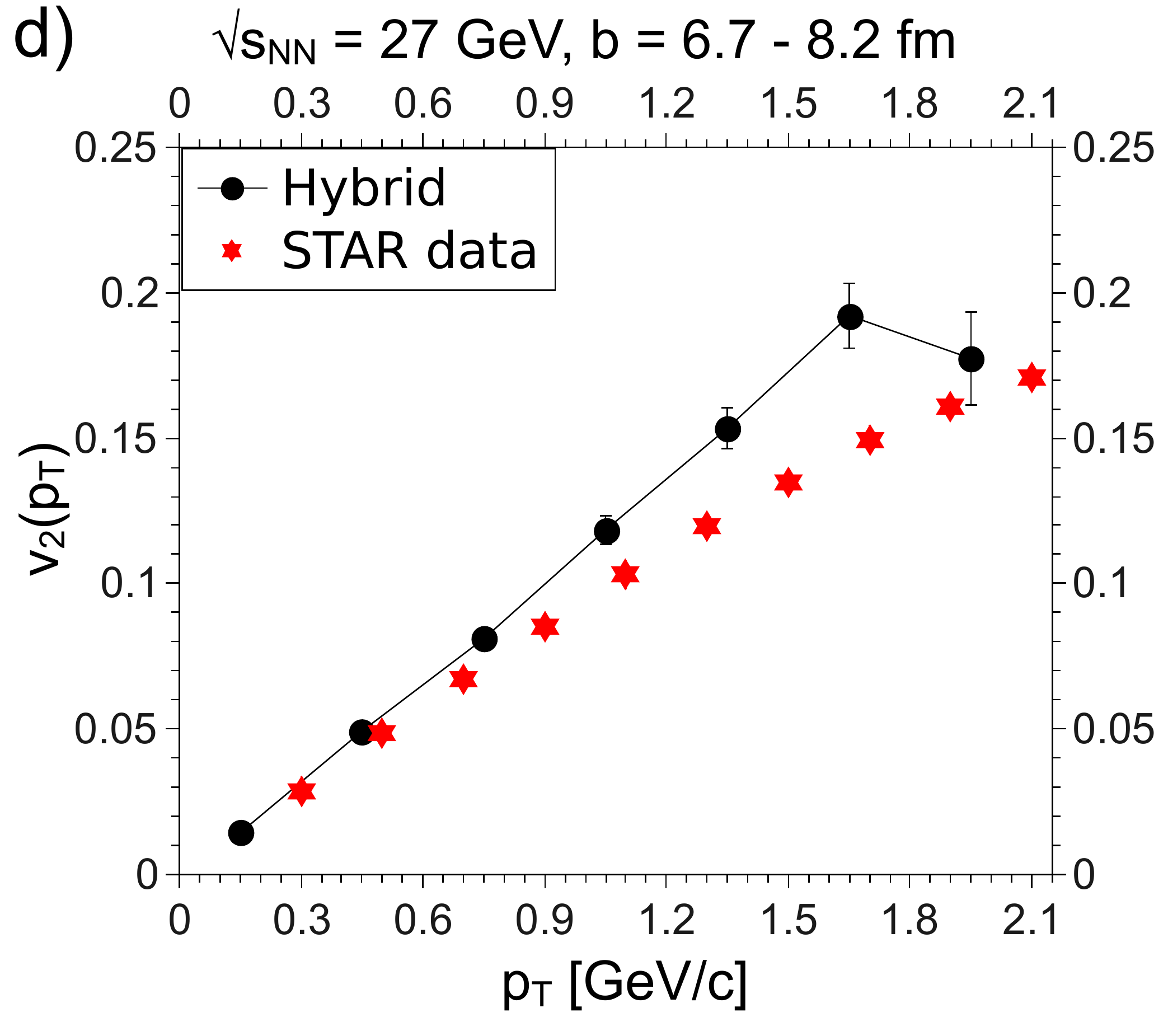}
\includegraphics[width=4.7cm]{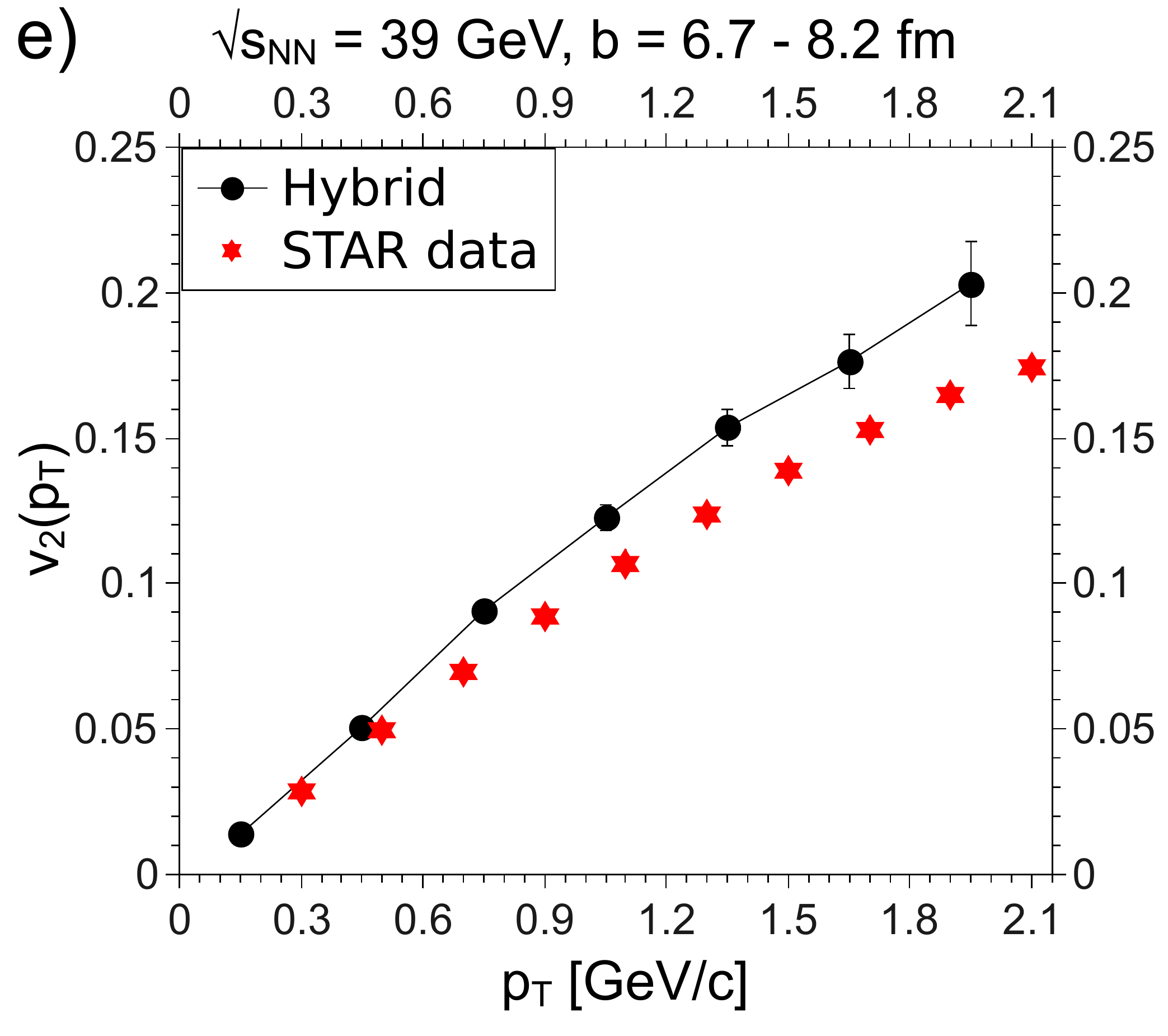}
\includegraphics[width=4.7cm]{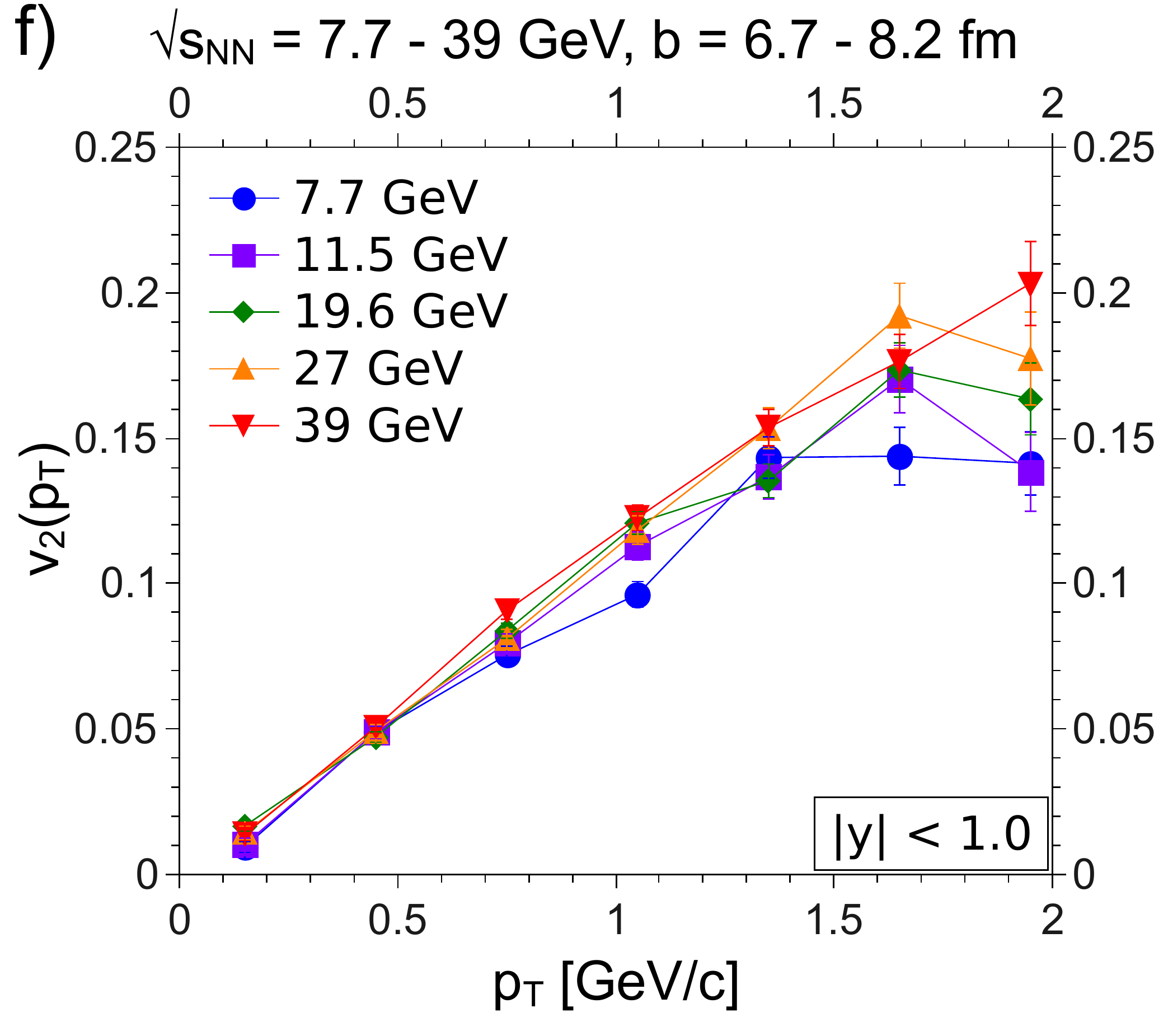}
\caption{(a-e) Differential $v_2$ at midrapidity $|y|<1.0$ for collision energies 
$\sqrt{s_{NN}}=7.7 - 39$ GeV in impact parameter range $b = 6.7-8.2$ fm, 
compared with the STAR data in (20-30)\% centrality \cite{Adamczyk:2012ku}. 
Panel (f): compilation of hybrid model $v_2(p_T)$ results from panels (a-e).}
\label{Figure_v2_pt_star}
\end{figure}

As seen in Figure~\ref{Figure_v2_pt_star}, $v_2(p_T)$ produced by the hybrid model 
systematically overshoots the data at all collision energies. This suggests the need for 
either adding viscous effects or stopping the hydrodynamical evolution earlier at a higher 
energy density. The dependence on the collision energy is non-existent, which is in accord 
with the STAR data, and is in this framework understood as the non-equilibrium hadron 
dynamics compensating for the shortened hydrodynamical evolution at lower $\sqrt{s_{NN}}$.

\subsection{Triangular flow}

\begin{figure}
\centering
\includegraphics[width=7.7cm]{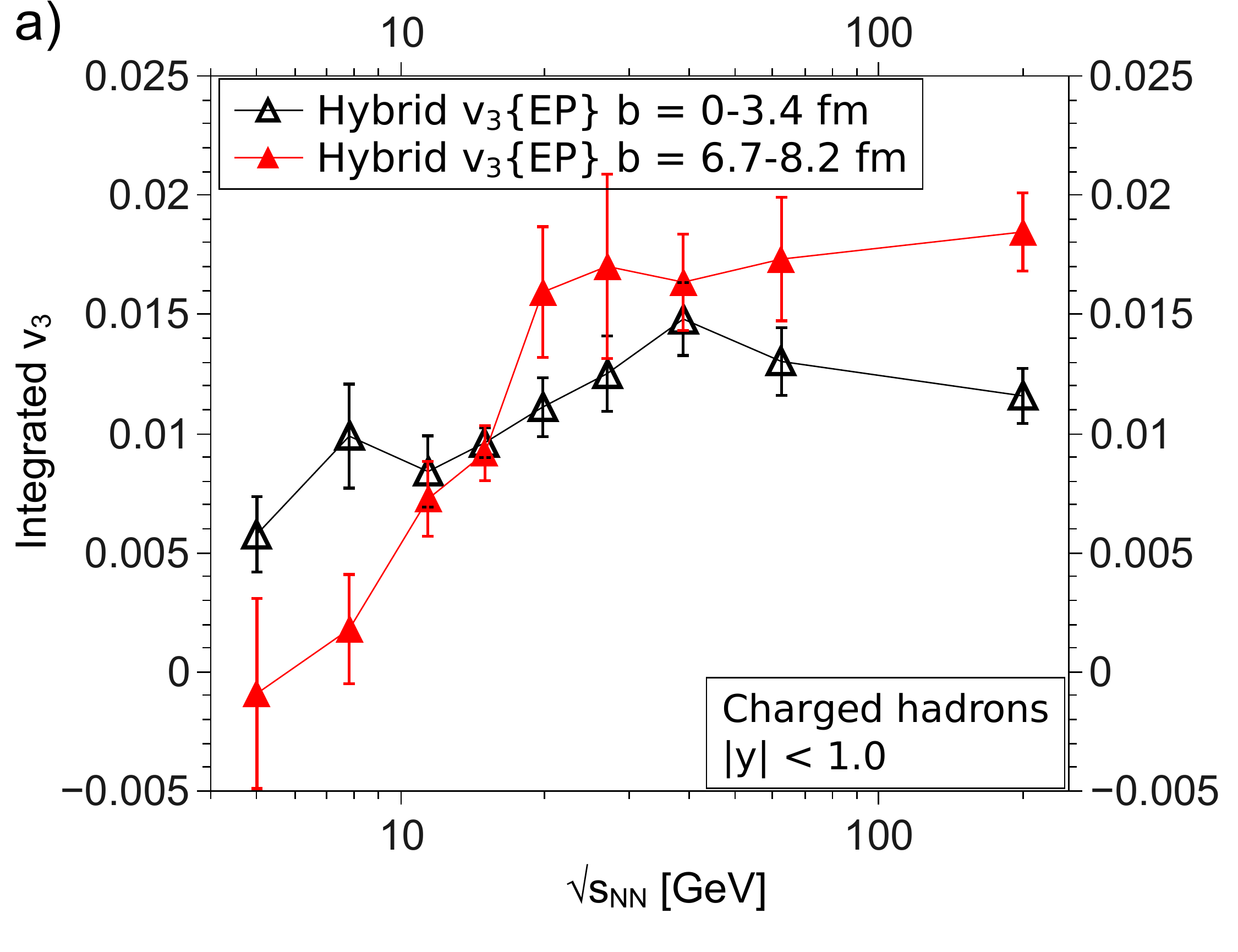}
\includegraphics[width=7.2cm]{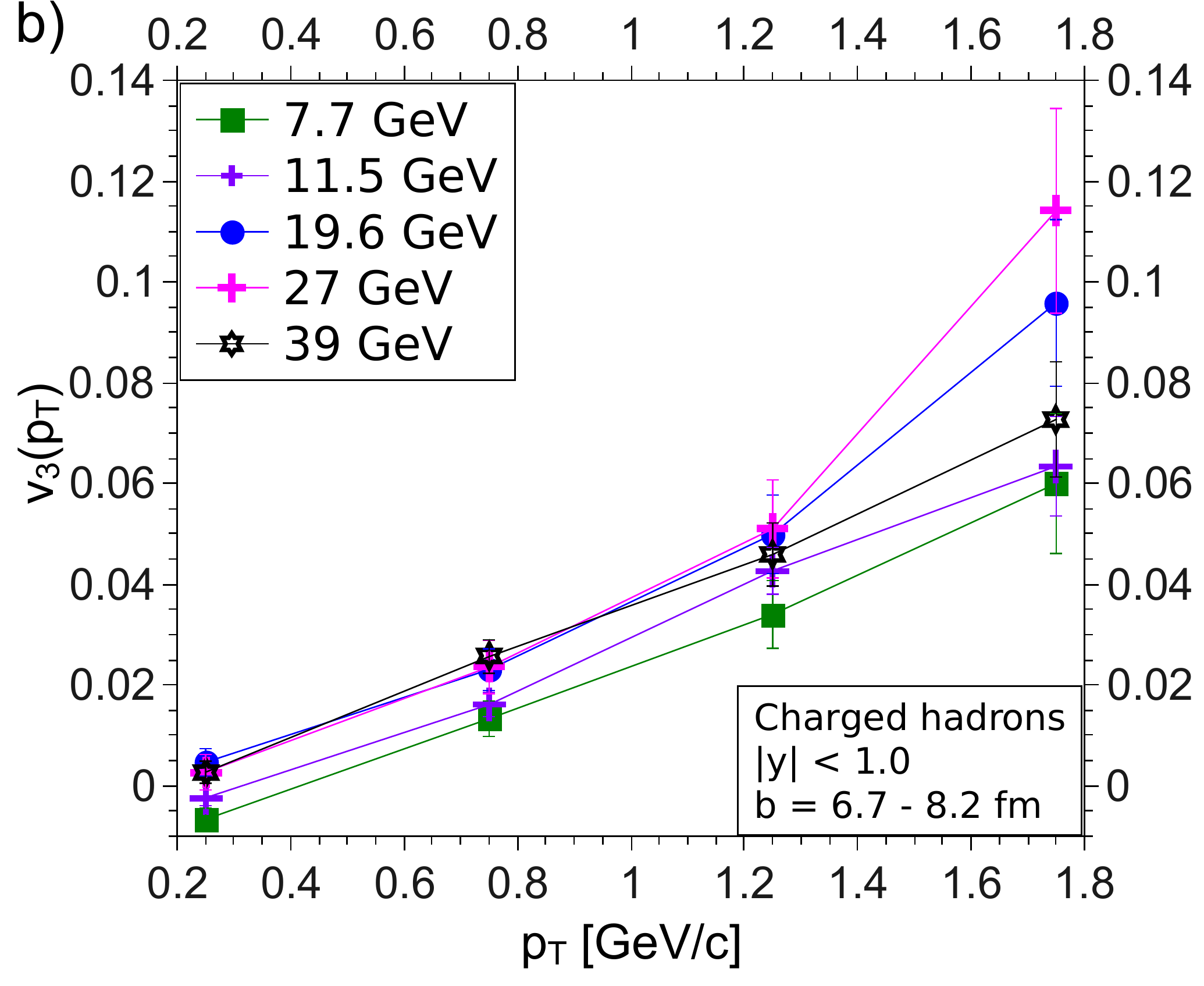}	
\caption{a) Integrated $v_3$ at midrapidity $|y|<1.0$ in central collisions ($b = 0-3.4$ fm, 
open triangles) 
and midcentral collisions ($b = 6.7-8.2$ fm, solid triangles). 
b) $v_3(p_T)$ in midcentral collisions.}
\label{Figure_v3}
\end{figure}

The triangular flow $v_3$ originates solely from the event-by-event variations in the 
initial configuration of the colliding nucleons, and is thus a good observable for measuring
the system sensitivity to the initial state fluctuations. In the most central collisions, 
integrated $v_3$ increases from 0.005 to above 0.01 with increasing collision energy 
(see Fig.~\ref{Figure_v3}a), whereas in midcentrality there is a rapid rise from $\approx 0$ 
at $\sqrt{s_{NN}} < 10$ GeV to the value of $\approx 0.015 - 0.02$ for 
$\sqrt{s_{NN}} \geq 19.6$ GeV. This behavior is reflected also on $v_3(p_T)$ 
in Fig.~\ref{Figure_v3}b. The energy dependence of $v_3$ is very similar to what was seen 
for the hydrodynamically produced $v_2$ in Figure~\ref{Figure_v2_phases}b, suggesting that 
in this case the transport part of the model is unable to compensate for the diminished 
hydro phase.

The magnitude of triangular flow at $\sqrt{s_{NN}}=200$ GeV is close to the measured value
for both centralities \cite{Adamczyk:2013waa}. However, the decrease to zero at low 
energies is not supported by the preliminary STAR data, where very little $\sqrt{s_{NN}}$
-dependence is seen below 30 GeV and the rise begins only at later energies 
\cite{Pandit:QM2012}. 

\subsection{Effect of initial geometry}

\begin{figure}
\centering
\includegraphics[width=7.2cm]{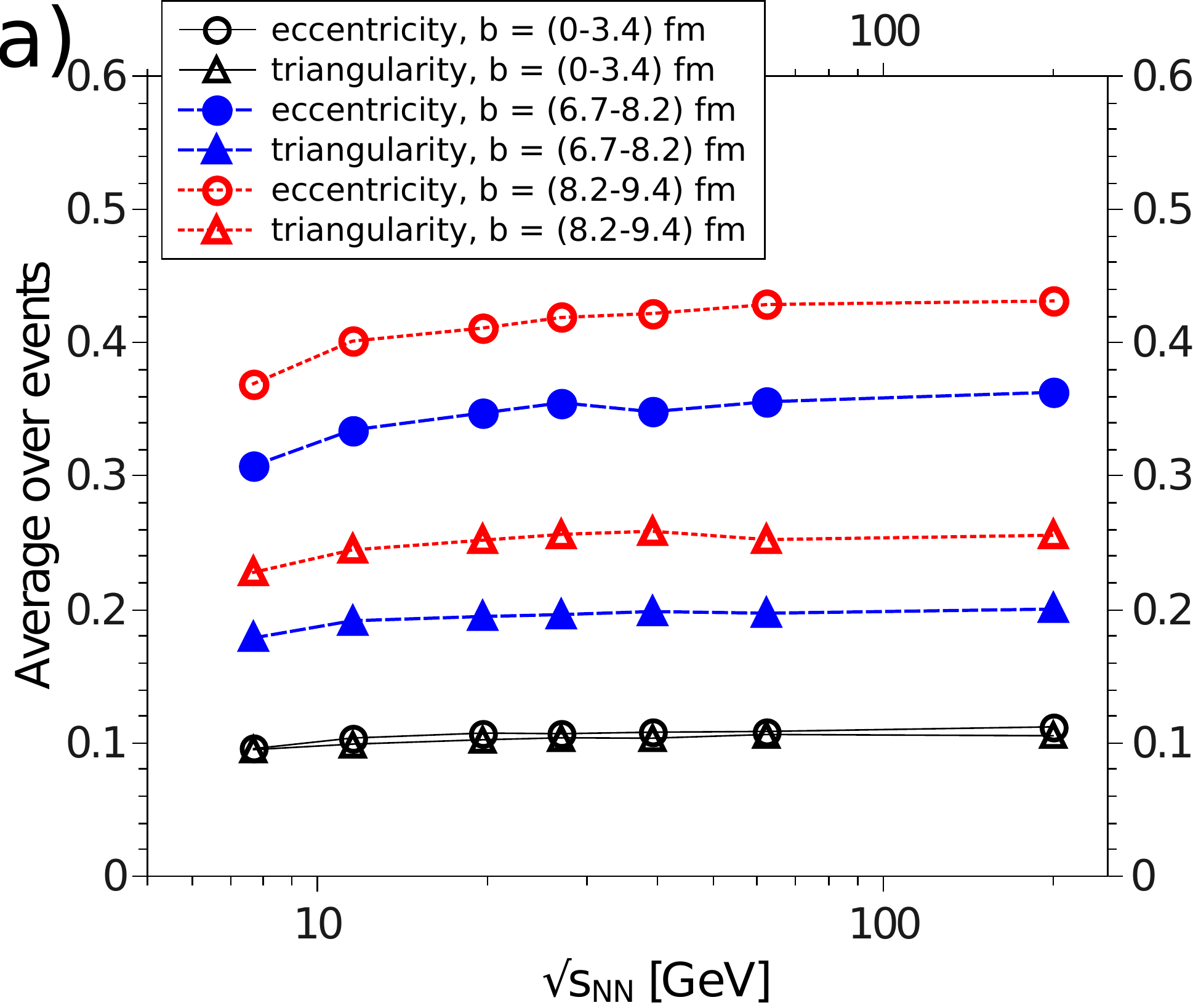}
\includegraphics[width=7.1cm]{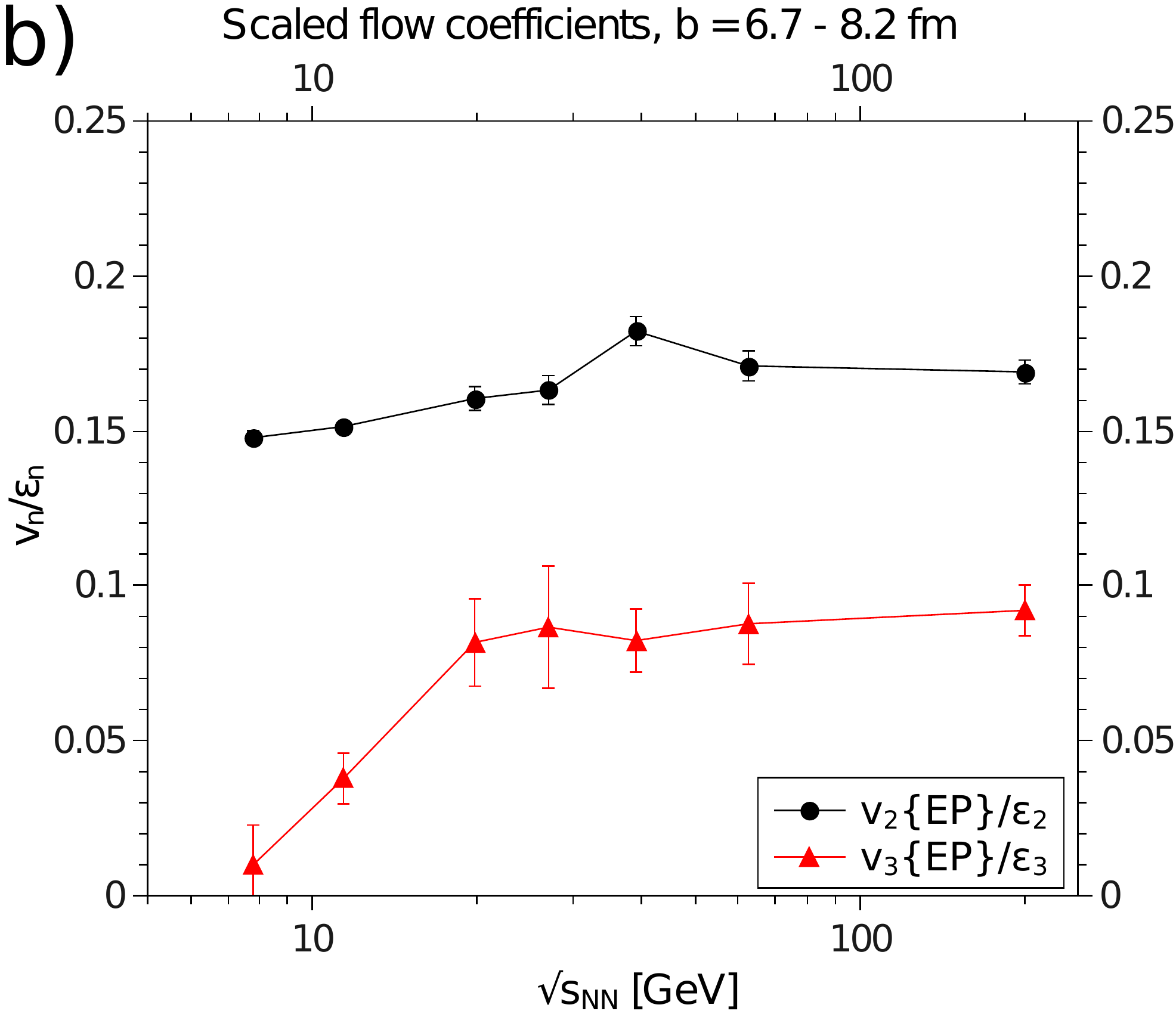}	
\caption{a) Average eccentricity $\langle \epsilon_2 \rangle$ (circles) and 
triangularity $\langle \epsilon_3 \rangle$ (triangles)
as a function of collision energy $\sqrt{s_{NN}}$, for impact parameter ranges
$b = 0-3.4$ fm (solid lines), $6.7-8.2$ fm (dashed lines) and $8.2-9.4$ fm (dotted lines).
b) Scaled flow coefficients $v_2 / \langle \epsilon_2 \rangle$ and 
$v_3 / \langle \epsilon_3 \rangle$ for $b = 6.7-8.2$ fm.}
\label{Figure_eccentricity}
\end{figure}

Figure~\ref{Figure_eccentricity}a illustrates the collision energy and centrality 
dependencies of the average initial state spatial eccentricity $\langle \epsilon_2 \rangle$ 
and triangularity $\langle \epsilon_3 \rangle$. The eccentricity and triangularity in 
an event are defined as in \cite{Schenke:2012hg} and calculated at the beginning of 
hydrodynamical evolution $t_{\textrm{start}}$.

In the most central collisions, both the average eccentricity and triangularity are similar 
in magnitude. The situation changes at mid-central collisions, where, due to the collision 
geometry, $\langle \epsilon_2 \rangle$ is clearly larger than $\langle \epsilon_3 \rangle$. 
There is only a weak dependence on the collision energy. This is not surprising, as 
neither the typical binary collision spatial distribution nor the inelastic nucleon-nucleon 
cross section $\sigma_{NN}$ are expected to change significantly within the examined energy 
range. What does change rapidly at lower collision energies is $t_{\textrm{start}}$, which 
drops from 3.22 fm at $\sqrt{s_{NN}}=7.7$ GeV to 1.23 fm at $\sqrt{s_{NN}}=19.6$ fm. This 
longer transport evolution would thus be the main reason for the systematic decrease of  
$\langle \epsilon_2 \rangle$ and $\langle \epsilon_3 \rangle$ at low 
energies in Figure~\ref{Figure_eccentricity}a.

In order to examine the system response to initial geometry, we scale $v_2$ and $v_3$ 
with $\langle \epsilon_2 \rangle$ and $\langle \epsilon_3 \rangle$, respectively. 
The result for $b = 6.7-8.2$ fm is shown in Figure~\ref{Figure_eccentricity}b. 
As the initial geometry displays little change over $\sqrt{s_{NN}}$, the result reflects 
what we already saw with the unscaled flow coefficients: the relation of the elliptic flow 
to the initial eccentricity remains almost constant for the whole collision energy range, 
while the $v_3$ response to the triangularity of the initial state reaches a constant value 
only after 19.6 GeV. This confirms that compared to hydrodynamics, the string / hadron 
transport dynamics are inefficient for transforming the initial state spatial fluctuations 
into the final state momentum anisotropy.

\section{Summary}

In this study, we have demonstrated that it is possible to reproduce the experimentally 
observed $v_2$ by utilizing a hybrid transport + hydrodynamics approach. In such a framework, 
it is seen that the hadron / string pre-equilibrium dynamics can compensate for the 
diminished hydrodynamical evolution for $v_2$ production at lower collision energies. For 
the triangular flow $v_3$ this is not true, and the system response to triangularity 
generated by the initial state fluctuations drops to near zero at the collision energies 
below 10 GeV.

However, while the values for the triangular flow $v_3$ at high collision energies 
quantitatively agree with the experimental results, there is a qualitative disagreement 
with the preliminary STAR data, which display non-zero $v_3$ at lower collision energies. 
As the transport dynamics have been proven ineffective for $v_3$ production in this 
investigation, this would suggest that the hydrodynamically behaving matter is manifested 
at the lower collision energies in greater extent than expected.

There are also issues with kaon production and $v_2(p_T)$ overestimating the data at 
higher $p_T$, which suggest that a slight re-tuning of the model parameters is required 
for the optimal agreement with the experimental data. These issues are revisited in the near 
future.

\section{Acknowledgements}

The authors acknowledge funding of the Helmholtz Young Investigator Group VH-NG-822. 
Computational resources have been provided by the Center for Scientific Computing (CSC) at 
the Goethe-University of Frankfurt.

\end{document}